\documentclass[twocolumn]{aastex62}

\newcommand{\Msun}{M_{\odot}} 
\newcommand{\Rsun}{R_{\odot}} 
\newcommand{\Lsun}{L_{\odot}}
\newcommand{\cmc}{\mathrm{cm}^{-3}}
\newcommand{\mdotsunyr}{M_\odot \mathrm{yr}^{-1}}
\usepackage{comment}

\graphicspath{{./}{figures/}}

\received{March 30, 2023}
\revised{May 30, 2023}
\accepted{May 30, 2023}

\submitjournal{ApJ}

\shorttitle{3D RHD simulations resolving protostellar interior}
\shortauthors{K. Kimura et al.}

\begin{document}

\title{3D Radiation-Hydrodynamic Simulations Resolving Interior of Rapidly Accreting Primordial Protostar}

\correspondingauthor{Kazutaka Kimura}
\email{kimura.k@astr.tohoku.ac.jp}

\author[0000-0001-8382-3966]{Kazutaka Kimura}
\affil{Center for Gravitational Physics and Quantum Information, \\
  Yukawa Institute for Theoretical Physics, Kyoto University, Kyoto, 606-8502, Japan}

\author[0000-0003-3127-5982]{Takashi Hosokawa}
\affiliation{Department of Physics, Graduate School of Science, Kyoto University, Sakyo, Kyoto 606-8502, Japan}
\author[0000-0001-7842-5488]{Kazuyuki Sugimura}
\affiliation{Department of Physics, Graduate School of Science, Kyoto University, Sakyo, Kyoto 606-8502, Japan}
\affiliation{The Hakubi Center for Advanced Research, Kyoto University, Sakyo, Kyoto 606-8501, Japan}
\affiliation{Faculty of Science, Hokkaido University, Sapporo, Hokkaido 060-0810, Japan}

\author[0000-0002-0547-3208]{Hajime Fukushima}
\affiliation{Center for Computational Sciences, University of Tsukuba, Ten-nodai, 1-1-1 Tsukuba, Ibaraki 305-8577, Japan}



%
%
\begin{abstract}
  Direct collapse of supermassive stars is a possible pathway to form supermassive black hole seeds at high redshifts. Whereas previous three-dimensional (3D) simulations demonstrate that supermassive stars form via rapid mass accretion, those resolving the stellar interior have been limited. We here report 3D radiation-hydrodynamic (RHD) simulations following the evolution of rapidly accreting protostars resolving the stellar interior. We use an adaptive mesh refinement code with our newly developed RHD solver employing an explicit M1 closure method. We follow the early evolution until the stellar mass reaches $\sim 10~\Msun$ from two different initial configurations of spherical and turbulent clouds. We demonstrate that, in both the cases, a swollen protostar whose radius is $100-1000~\Rsun$ appears, as predicted by the stellar evolution calculations. Its effective temperature remains a few thousand Kelvin, and the radiative feedback by ionizing photons is too weak to disturb the accretion flow up to the epoch examined in this work. In the turbulent case, the protostar rotates rapidly at more than 0.4 times the Keplerian velocity owing to the angular momentum provided by the initial turbulence. The protostar approximates an oblate spheroid, and its equatorial radius is more than twice the polar radius. Our results suggest that we need to consider the rapid stellar rotation to elucidate the realistic 3D protostellar evolution in the supermassive star formation.
\end{abstract}



%
%
\section{Introduction} \label{Sec:Intro}
\par
Many supermassive black holes (SMBHs) exceeding $10^9~\Msun$ exist in the high-redshift universe ($z\ge6$), and their possible origins have been intensively studied \citep[see][for reviews]
{Volonteri_2010,Woods_et_al_2019,Inayoshi_et_al_2020}.
One possible pathway to form such SMBHs is the direct collapse (DC) scenario, which starts with the primordial gas cloud collapsing via H atomic cooling \citep{Bromm_and_Loeb_2003}.
In this case, the cloud maintains high temperatures ($\gtrsim 3000$~K), and protostars grow in mass with high accretion rates of $\sim 0.1\mathrm{-}1~\mdotsunyr$ \citep[e.g.][]{Omukai_2001,Latif_et_al_2013,Chon_et_al_2018}. If such rapid accretion continues for $\sim 1$~Myr, protostars evolve into supermassive stars (SMSs) with masses of $\sim 10^5\mathrm{-}10^6~\Msun$.
Finally, they collapse into BHs with the same masses due to general relativistic instability \citep{Zeldovich_and_Novikov_1971,Shapiro_and_Teukolsky_1983,Shibata_and_Shapiro_2002}. Such BH seeds can grow into the high-redshift SMBHs even with sub-Eddington accretion rates.
\par
Evolution of the protostellar structure is crucial in this scenario. If a protostar emits a copious amount of ionizing photons,
the strong radiative feedback may hinder the mass accretion \citep[e.g.][]{McKee_and_Tan_2008,Hosokawa_et_al_2011,Sugimura_et_al_2020} before the formation of a SMS. Stellar evolution calculations predict that rapidly accreting massive protostars should have very large radii and low effective temperatures \citep{Hosokawa_et_al_2012, Hosokawa_et_al_2013,Schleicher_et_al_2013,Umeda_et_al_2016,Haemmerle_et_al_2018a}. As a result, their ionizing emissivities remain very low, even if the luminosities approach the Eddington limit.
\par
One-dimensional (1D) stellar evolution calculations are a major methodology in studying the protostellar evolution.
However, 1D stellar models suffer from limitations. For instance, it normally assumes the hydrostatic balance for the stellar materials and steady accretion flow for the surroundings. Furthermore, realistic three-dimensional (3D) structure, such as stellar rotation provided by angular momentum carried by the accretion flow, is difficult to be modeled. The protostellar rotation should affect the evolution as it induces efficient chemical mixing in the interior. With fast rotation, moreover, the centrifugal force may limit mass accretion through a surrounding disk via the so-called $\Omega\Gamma$-limit \citep{Maeder_and_Meynet_2000,Lee+Yoon2016,Takahashi_and_Omukai_2017,Haemmerle_et_al_2018b}. The actual protostellar evolution and resulting feedback may differ from predictions by previous 1D calculations.
\par
To elucidate the realistic 3D protostellar evolution, we need to perform three-dimensional radiation hydrodynamic (RHD) simulations.
\citet{Luo_et_al_2018} report 3D RHD simulations following an early evolution in the SMS formation with the flux-limited diffusion approximation, resolving the stellar interior structure. They show that a protostar is dissolved by the growing pressure, which differs from the previous 1D calculations. We need further study to address how generally such evolution occurs and what 3D RHD effects may cause it.
\par
In this work, we perform 3D simulations using a newly developed RHD solver with an explicit M1 closure scheme, resolving the protostellar interior. We present simulations starting from two initial configurations: spherical and turbulent clouds. We study the idealized spherical case for comparisons to the 1D calculations and the turbulent case for revealing realistic protostellar evolution.
We follow the early evolution of the SMS formation until a protostar accretes the gas of $\sim 10~\Msun$.
\par
The organization of this paper is as follows. We describe our numerical method in Section \ref{Sec:Method}.
In Section \ref{Sec:Results_Sphe} and \ref{Sec:Results_Turb}, we show our simulation results for the spherical and turbulent cases respectively.
In Section \ref{Sec:Discussions}, we discuss the implication of our results.
Finally we give summary in Section \ref{Sec:Summary}.

%
%
\section{Numerical Method} \label{Sec:Method}
\subsection{Radiation Hydrodynamics Code}
\par
We use the self-gravitational magneto-hydrodynamics code with adaptive mesh refinement, SFUMATO-RT \citep{Matsumoto_2007, Matsumoto_et_ak_2015, Sugimura_et_al_2020}.
SFUMATO-RT includes the primordial gas chemistry module, which applies only to a regime where the density is lower than $10^{13}~\cmc$. We add chemical reactions that are relevant for a high-density medium \citep[see also][]{Sadanari_et_al_2021}.
Moreover, we implement the newly developed radiation solver with an explicit M1 closure scheme applicable even for the optically thick regime \citep{Pomraning_1969,Kershaw_1976,Levermore_1984,Rosdahl_and_Teyssier_2015,Fukushima_and_Yajima_2021}.
We consider the radiation force with this solver.
We compute the local opacity consistently with the non-equilibrium chemistry, following the method developed in \citet{Matsukoba_et_al_2019}.
We do not use the ray-tracing radiation solver implemented by \citet{Sugimura_et_al_2020} in this work. Our radiation solver will be separately described in detail in a forthcoming paper (K. Kimura et al., in preparation).
\par
As for the chemistry network, we solve the non-equilibrium reactions among six components, $\mathrm{H}$, $\mathrm{H}_2$, $\mathrm{H}^+$, $\mathrm{H}^-$, $\mathrm{H}_2^+$, and $\mathrm{e}$ as in \citet{Sugimura_et_al_2020}.
We set the fractional abundance of He to be $y_\mathrm{He} = 9.722\times10^{-2}$
We have implemented chemical reactions effective in the dense medium with $n_\mathrm{H} \gtrsim 10^{13}~\cmc$ following \citet{Omukai_2001}. To follow the coupled evolution of the gas and radiation in the optically thick medium, we incorporate the photon emission and absorption processes into our chemical network. We simultaneously update the gas temperature, chemical abundances, and radiation fields using the implicit method, to solve their evolution consistently.
\par
We develop a new explicit M1 closure scheme based on \citet{Rosdahl_and_Teyssier_2015}.
Their method is applicable to both the optically thin and thick regions, particularly for cases where the optical depth over a single grid cell greatly exceeds unity. We have modified their scheme to handle a situation where the optical thickness abruptly changes across a few cells. This occurs near the protostellar surface in our simulations.
\par
We consider two frequency bins in the radiation solver: low energy ($h\nu<13.6~\mathrm{eV}$) and extreme-ultraviolet (EUV; $h\nu>13.6~\mathrm{eV}$).
In the SMS formation, the emission of the low-energy radiation is crucial for the cooling processes of the collapsing gas.
Meanwhile, the protostar emits a large amount of EUV photons, which ionize the surrounding gas and suppress the gas accretion, if the protostellar effective temperature rises to $\sim10^5$~K.
Taking these two frequency bins is thus necessary to study the SMS formation.
Furthermore, for each frequency bin, we separately solve the photon energy and number densities, whose ratio provides the photon mean energy. We locally reconstruct a consistent spectral distribution within each bin, assuming the Planck distribution with the temperature corresponding to the mean energy. This method improves our approximate treatment with only two bins, allowing us to follow the radiation physics such as heating by absorption accurately. Note that the luminosity shown in Section~\ref{Sec:Results} is the total of low-energy and EUV components.
\subsection{Initial Conditions and Settings}
\par
One initial condition we consider is the ``spherical'' case, where we start with an isothermal Bonner-Ebert sphere \citep{Bonnor_1956} with the central density $10^9~\cmc$ and temperature $4.8 \times 10^3~\mathrm{K}$. These values are realized on the SMS-evolution path given by the one-zone model of the cloud collapse \citep{Omukai_2001}. We increase the density profile by a factor of $1.6$ to ensure the collapse.
\par
The other initial condition is the ``turbulent'' case, where we start from the same cloud as above but with additional turbulent velocity fields. We assume the power spectrum of the turbulence obeying Larson's law, $P(k) \propto k^{-4} $, where $k$ is the wavenumber \citep{Larson_1981}. We set its Mach number as unity, considering recent studies showing that the turbulence grows up to this level during the collapse stage \citep{Federrath_et_al_2011, Higashi_et_al_2021}.
\par
We set the side length of the computational domain as $0.09\,\mathrm{pc}$, 3.3 times larger than the radius of the Bonner-Ebert sphere.
The cell number of the base grid is 40 in each direction, and we adaptively refine the grids to resolve the Jeans length with 8 cells.
The corresponding minimum cell size is $3.4 \times 10^{-3}\,\mathrm{AU}$ in the spherical case and $6.9 \times 10^{-2}\,\mathrm{AU}$ in the turbulent case.
We prohibit the grid de-refinement because it artificially increases the entropy in the stellar interior.
\par
In our explicit implementation of the M1 closure scheme, we adopt the reduced speed of light approximation to increase time steps restricted by the Courant-Friedrichs-Lewy condition \citep{Gnedin_and_Abel_2001,Skinner_and_Ostriker_2013}. We set the reduced light speed $\tilde{c}$ as $10^{-3}c$, where $c$ is the speed of light.
\par
The appearance of a standing accretion shock marks the epoch of the protostellar birth \citep{Omukai_and_Nishi_1998}. We define the stellar radius as the position at which the radial velocity falls below the sound speed and the photospheric radius as where the optical depth measured from the outside exceeds unity. The photospheric radius often becomes larger than the stellar radius, and the region between them is called a radiative precursor \citep{Stahler_et_al_1980}. We regard the total mass inside the stellar surface as the protostellar mass.
\par
In both the cases, we follow the evolution for $\sim 10$~yr after the birth of an embryonic protostar. The protostar accretes the gas of $\sim 10~\Msun$ by the end of the simulations with the mean accretion rate of $\sim 1~\mdotsunyr$.
The spherical and turbulent simulation runs have required 2 weeks with 896 cores and 2 months with 2000 cores, respectively.

%
%
\section{Results} \label{Sec:Results}
\par
\begin{figure*}
  \begin{center}
    \includegraphics[width=\linewidth]{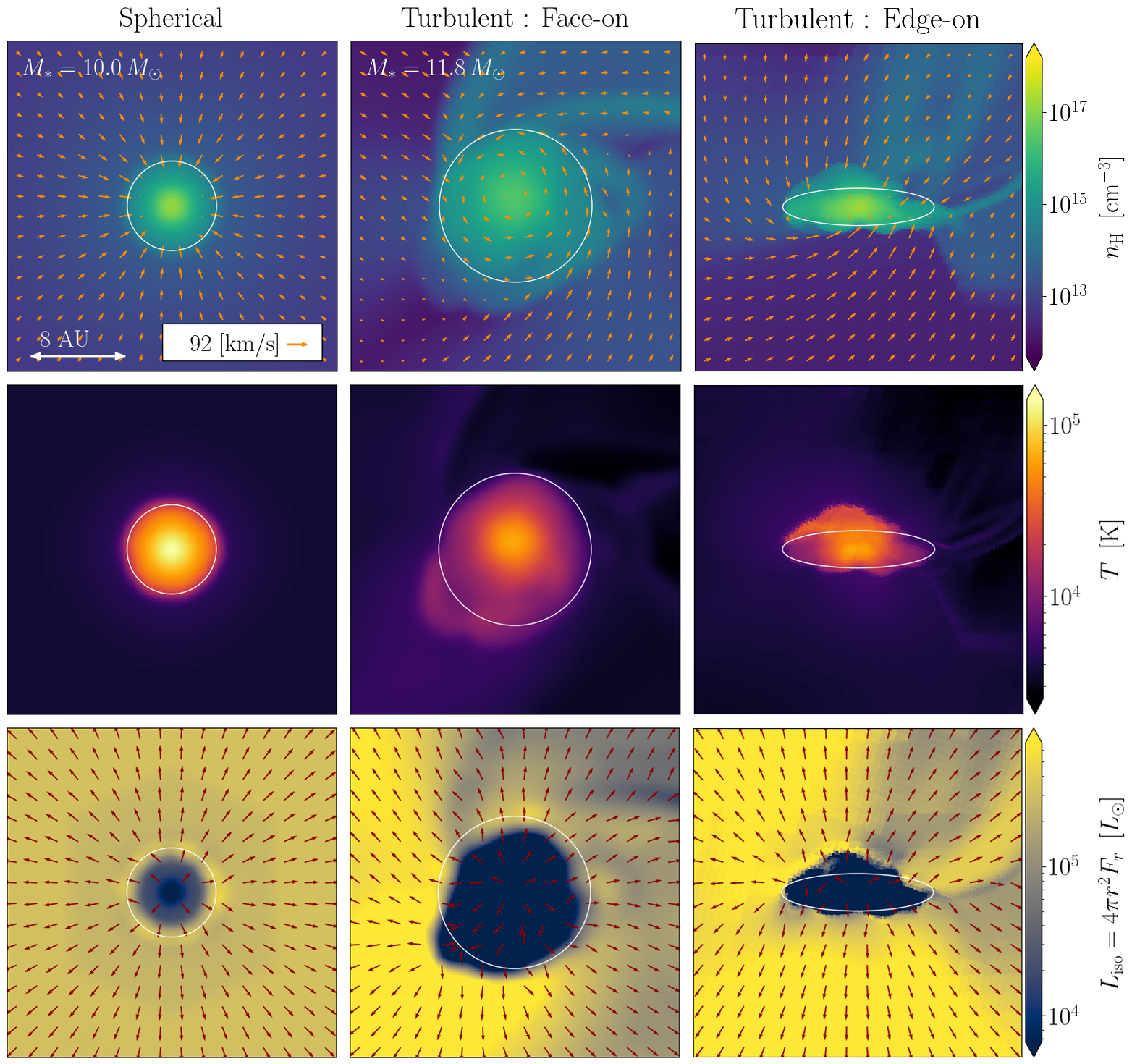}
    \caption{Snapshots at the end of the simulation.
      The left column corresponds to the spherical case. The middle and right columns show the face-on and edge-on slices in the turbulent case.
      The top, middle, and bottom panels display the distributions of the density, temperature, and isotropic luminosity $L_{\rm iso} \equiv 4\pi r^2 F_r$, where $F_r$ is the local outward radiation flux.
      The isotropic luminosity represents the flux intensity corrected for the geometrical dilution. In the top panel, arrows represent velocity distributions.
      In the bottom panel, arrows only represent directions of the radiation flux.
      The white ellipses approximately delineate the protostellar surfaces.
    }
    \label{fig:Snapshot}
  \end{center}
\end{figure*}

Figure~\ref{fig:Snapshot} shows the final snapshots for the spherical and turbulent cases. An accreting protostar with $M_* \sim 10~\Msun$ appears for both cases, and it has a large radius of $\sim 10^3~\Rsun$. Nonetheless, there are significant differences between these cases.
The turbulent case especially shows that the protostar rotates and has a flattened shape. We describe the protostellar structure and evolution for each case in detail below.
\subsection{Spherical Case} \label{Sec:Results_Sphe}
\par
\begin{figure}
  \begin{center}
    \includegraphics[width=\linewidth]{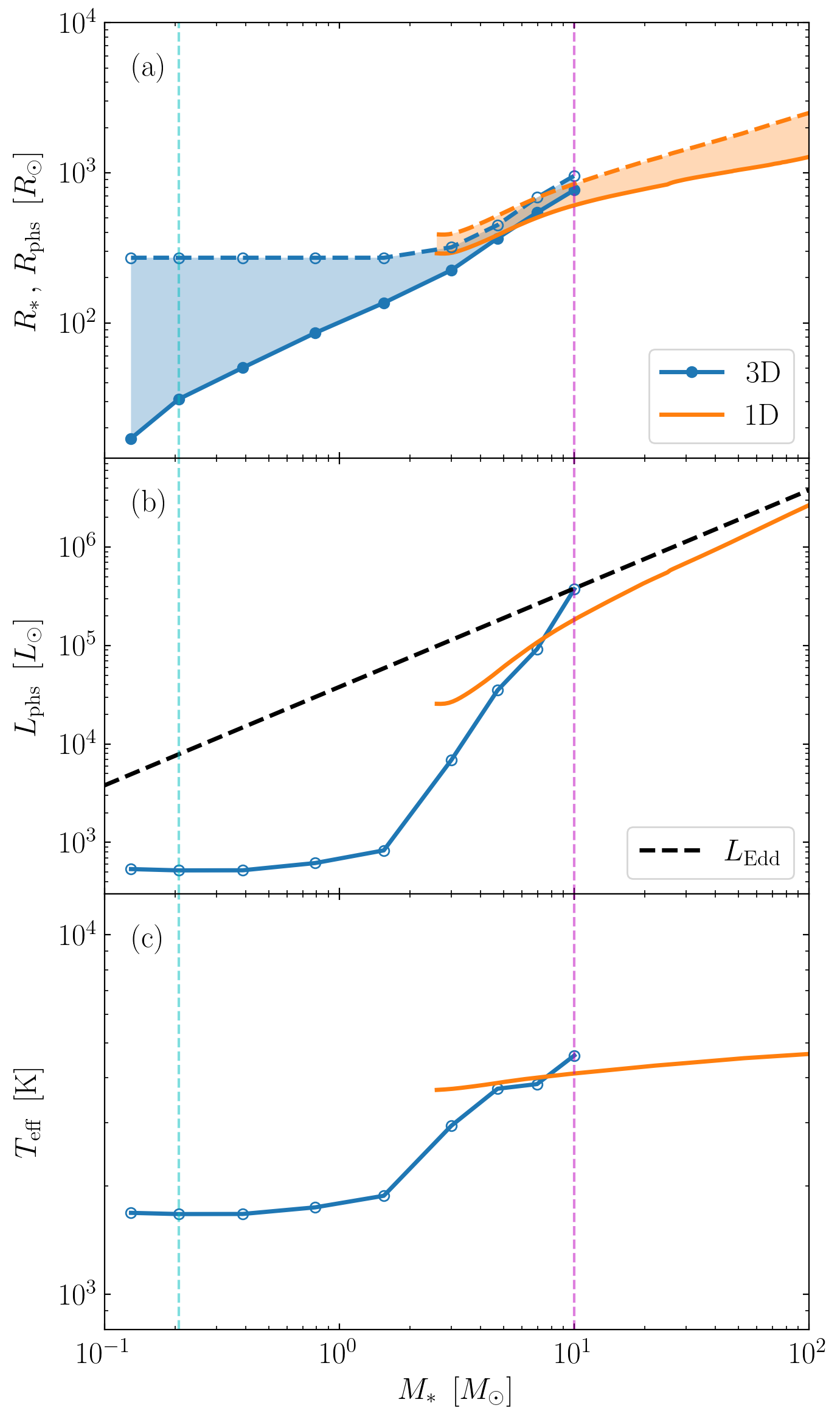}
    \caption{Stellar evolution in the spherical case. Panels show the evolution of (a) protostellar radius $R_*$ and photospheric radius $R_\mathrm{phs}$, (b) photospheric luminosity $L_\mathrm{phs}$,
      and (c) effective temperature $T_\mathrm{eff} \equiv (L_\mathrm{phs}/(4\pi R_\mathrm{phs}^2 \sigma_\mathrm{SB}))^{1/4}$, where $\sigma_\mathrm{SB}$ is the Stefan-Boltzmann constant, against the stellar mass $M_*$.
      In all the panels, the blue and orange colors represent our 3D simulation and the 1D stellar evolution calculation at the constant accretion rate $1~\mdotsunyr$ taken from \citet{Hosokawa_et_al_2012}.
      The open and filled circles denote the actual values evaluated from the simulation data. In panel (a), the solid and dashed lines represent the protostellar and photospheric radii, respectively. The shaded regions correspond to the radiative precursor. In panel (b), the dashed line indicates the Eddington luminosity defined with the Thomson scattering cross section. The vertical cyan and magenta lines mark the epochs whose radial profiles are presented in Figure \ref{fig:Sphe_profile}.
    }
    \label{fig:Sphe_StellarEvol}
  \end{center}
\end{figure}

In this case, the spherical symmetry is kept throughout the evolution as suggested by Figure~\ref{fig:Snapshot}. Figure~\ref{fig:Sphe_StellarEvol} shows the protostellar evolution, where a protostar first appears at $M_* \sim 0.1~\Msun$. The top panel shows at this epoch the stellar radius is $\simeq 10~\Rsun$, much smaller than the photospheric radius $\simeq 300~\Rsun$. The protostellar radius is comparable to the Jeans length when the collapsing gas becomes adiabatic at $n_\mathrm{H} \sim 10^{20}~\cmc$ \citep{Omukai_2001}, and the photospheric radius is determined by the envelope structure created during the collapse.
As the accretion proceeds, only the protostellar radius grows and approaches the photospheric radius. These radii increase in tandem for $M_* \gtrsim 3~\Msun$, after the stellar radius exceeds $\sim 100~\Rsun$. The protostellar and photospheric radii at this stage agree with the 1D stellar evolution calculation taken from \citet{Hosokawa_et_al_2012}, for which the steady spherical accretion at a constant rate $1~\mdotsunyr$ was assumed.
Note that the 1D results are not available for $M_* \lesssim 2.5~\Msun$, because of numerical difficulties in constructing models. \citet{Hosokawa_et_al_2012} adopt an arbitrary $M_* \simeq 2.5~\Msun$ initial model with the radius $R_* \simeq 298.4~\Rsun$ instead. In contrast, our 3D simulation consistently follows the early evolution from the birth of the protostar.
\par
The middle panel in Figure~\ref{fig:Sphe_StellarEvol} shows the evolution of the luminosity at the photosphere.
In the early stage for $M_* \lesssim 2~\Msun$, the radiative precursor near the photosphere has the same structure as that created during the collapse. As a result, the luminosity remains almost constant at $\simeq 5 \times 10^2~\Lsun$.
Later for $M_* \gtrsim 2~\Msun$, the photosphere expands, and the luminosity increases to reach the Eddington values. However, the radiative feedback is ineffective for the accreting gas because it is neutral and the opacity is lower than that of Thomson scattering.
The lower panel shows the evolution of the effective temperature $T_\mathrm{eff}$. We see that $T_\mathrm{eff}$ takes low values of a few $\times 10^3$~K in spite of the large luminosities because of the large radii. These features agree with the 1D calculation.
\begin{figure}
  \begin{center}
    \includegraphics[width=\linewidth]{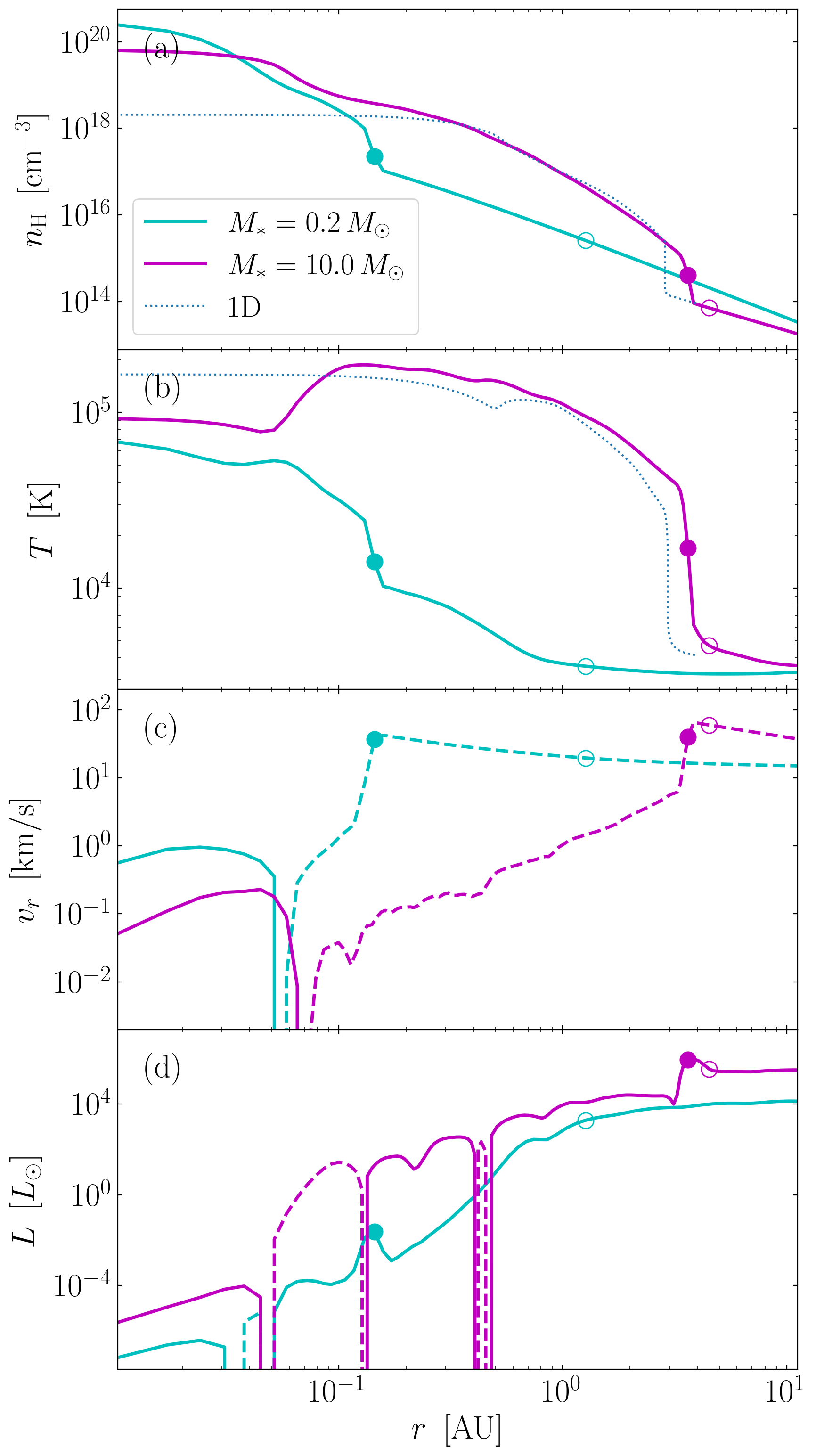}
    \caption{Spherically averaged radial profiles of physical quantities in the spherical case. The colors represent different epochs when the stellar mass is $0.2~\Msun$ (cyan) and $10.0~\Msun$ (magenta). The filled and open circles denote the protostellar and photospheric radii (see the top panel in Fig.~\ref{fig:Sphe_StellarEvol}). Panels show the distributions of (a) gas number density, (b) temperature, (c) radial velocity, and (d) luminosity against the distance from the stellar center. In panels (c) and (d), the solid and dashed lines correspond to the outward and inward flux, respectively.
      In panels (a) and (b), the dotted lines represent the 1D stellar evolution calculations by \citet{Hosokawa_et_al_2012}, the profiles when the stellar mass reaches $10~\Msun$ with the constant accretion rate of $1~\mdotsunyr$.
    }
    \label{fig:Sphe_profile}
  \end{center}
\end{figure}
\par
Figure~\ref{fig:Sphe_profile} shows the spherically averaged radial profiles of physical quantities when the stellar mass is $0.2~\Msun$ and $10.0~\Msun$. Panels (a)--(c) show the presence of an accretion shock at the stellar surface, where the infall velocity drops and both the density and temperature abruptly rise instead.
Throughout the evolution, the stellar interior is fully radiative, or convectively stable, as expected by the 1D calculation for the early ``adiabatic accretion'' stage \citep[e.g.][]{Hosokawa_and_Omukai_2009}.
\par
For comparison, we also show the profiles of 1D stellar evolution calculations by \citet{Hosokawa_et_al_2012}, the snapshots at the epoch when the stellar mass is $\simeq 10~\Msun$ with the constant accretion rate of $1~\mdotsunyr$.
Our results agree well with the 1D calculations except for the innermost part of $r \lesssim 0.1$~AU. Note that the 1D calculations do not solve the initial core formation but assume an arbitrary initial model with $2.5~\Msun$. In contrast, our 3D simulations follow the protostellar accretion consistently from the embryonic protostar with $\sim 0.1~\Msun$. This causes the different structures in the deep stellar interior at $M_* \simeq 10~\Msun$.
\par
Panel (d) shows that the luminosity in the deep stellar interior is small due to slow diffusion. Each profile has the local maximum near the surface owing to the large temperature gradient created by the shock heating.
When $M_*=0.2~\Msun$, the luminosity increases outward in the radiative precursor since the temperature decreases and the $\mathrm{H}^-$ opacity consequently decreases outward. Moreover, at this time, the luminosity slightly increases further outward from the photosphere.
This is because the $\mathrm{H}^-$ free-bound emission, the dominant cooling process for the envelope gas, produces low energy ($h\nu < 13.6$~eV) radiation and contributes to the luminosity.
\subsection{Turbulent Case} \label{Sec:Results_Turb}
\begin{figure}
  \begin{center}
    \includegraphics[width=\linewidth]{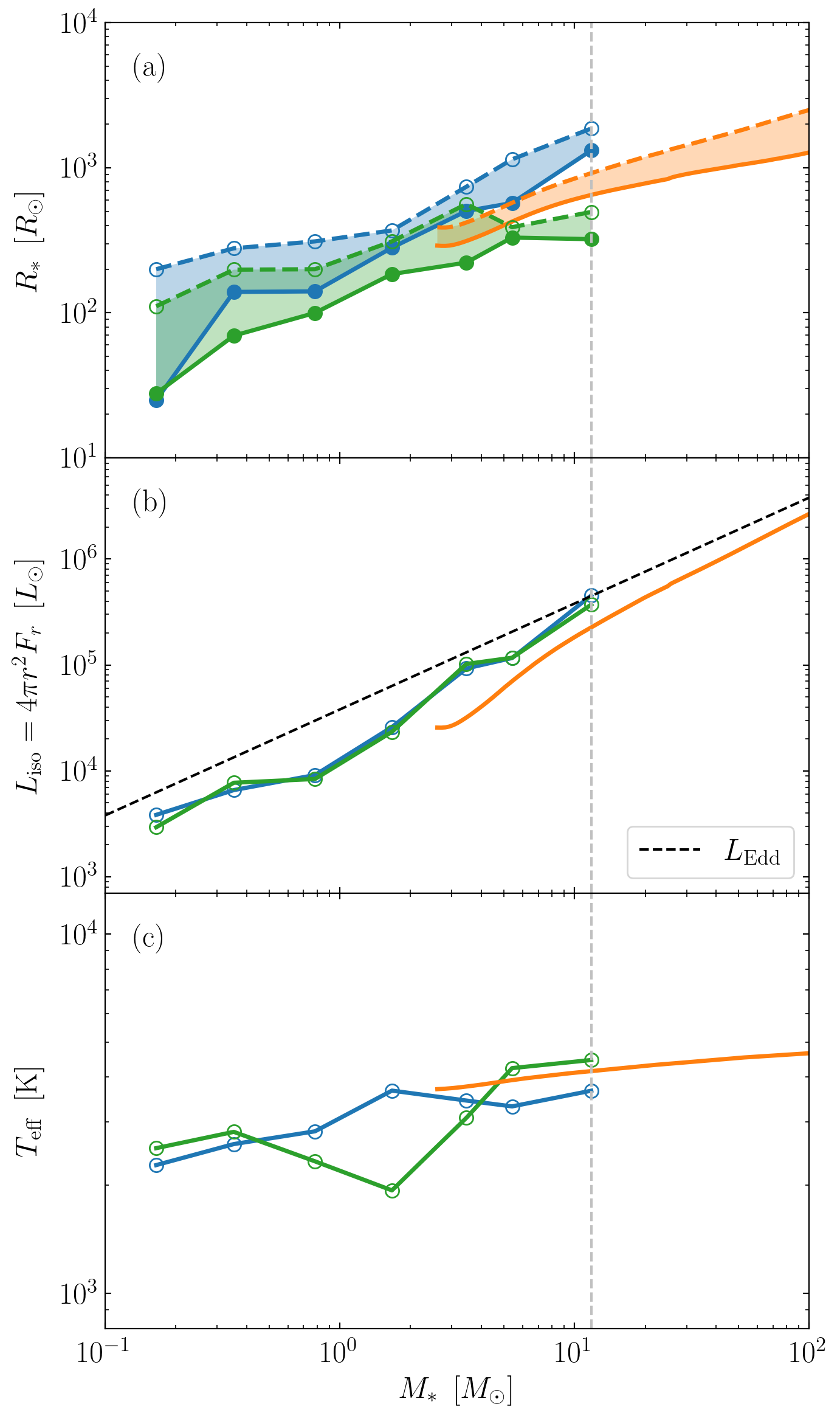}
    \caption{Stellar evolution in the turbulent case. Panels (a) and (c) show the same quantities as in Figure~\ref{fig:Sphe_StellarEvol}, and Panel (b) presents the isotropic luminosity.
      The blue and green circles show the quantities in the equatorial and polar directions.
      In panel (a), the equatorial radius (blue) indicates that averaged over the equatorial plane. The polar radius (green) is the minimum value on the polar axis, i.e., whether in the positive or negative directions. We use this to avoid a temporal dense structure created by the turbulence and resulting spurious large radius. Note that these equatorial and polar radii define the ellipses shown in Figure~\ref{fig:Snapshot}. In panels (b) and (c), the equatorial and polar values correspond to those averaged over the region within $\pi/4$ radians from the equatorial plane and polar axis.
      The vertical dashed line indicates the epoch whose snapshots are shown in Figures \ref{fig:Snapshot} and \ref{fig:Turb_profile}.
    }
    \label{fig:Turb_StellarEvol}
  \end{center}
\end{figure}
\begin{figure}
  \begin{center}
    \includegraphics[width=\linewidth]{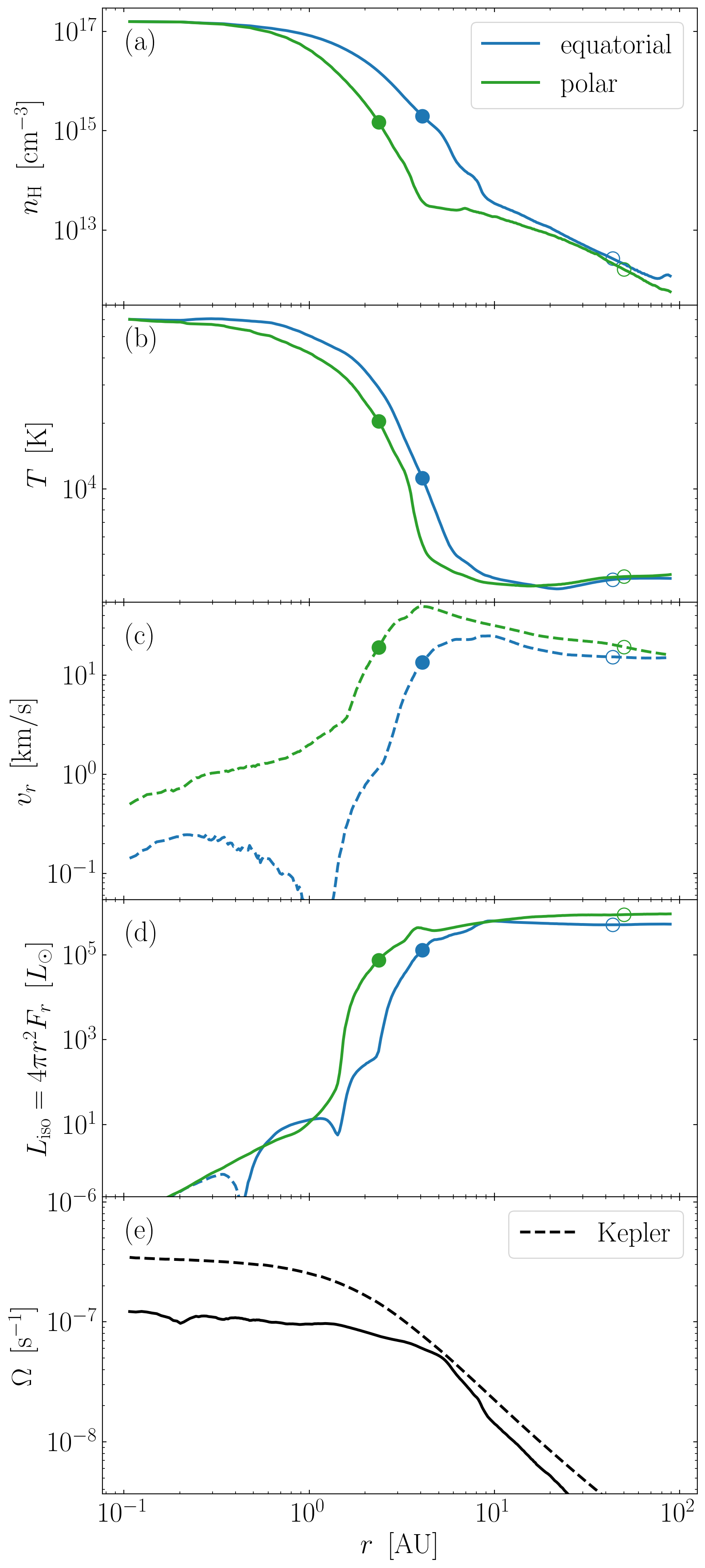}
    \caption{Radial profiles of physical quantities at the end of the turbulent case. Panels (a)--(d) show the same quantities as in Figure \ref{fig:Sphe_profile}. The colors represent the averaged radial profiles over the region within $\pi$/4 radians from the equatorial plane (blue) and polar axis (green). The filled and open circles denote the protostellar and photospheric radii. Panel (e) shows the spherically averaged angular velocities compared to the Keplerian values.
    }
    \label{fig:Turb_profile}
  \end{center}
\end{figure}
\par
We show the protostellar evolution in Figure \ref{fig:Turb_StellarEvol}.
As shown in the middle and right columns of Figure~\ref{fig:Snapshot}, the protostar has a complicated shape deviated from the spherical symmetry. Here, we show the evolution in equatorial and polar directions separately. Panel (a) illustrates that the stellar and photospheric radii increase with the increasing stellar mass. The stellar and photospheric radii are of the same order of magnitude as the 1D calculation for $M_* \gtrsim 3~\Msun$.
Meanwhile, the equatorial radii are more than twice as large as the polar ones because of the rotation. The protostar rotates so fast that the centrifugal force is comparable to the pressure gradient force near the surface in the equatorial direction, affecting the equilibrium shape, as shown below.
\par
Panel (b) presents that, at the end of the simulation, the luminosity is about the same as the 1D calculation and near the Eddington value.
At this time, the radiative feedback is ineffective because the surrounding gas is neutral as in the spherical case. Whereas the radiative flux is anisotropic depending on the accreting gas structure (Figure~\ref{fig:Snapshot}), the averaged luminosity over the equatorial and polar regions is almost the same.
Panel (c) shows that the effective temperature is always a few thousand Kelvin and agrees with the 1D calculation. The resulting UV feedback is too weak to disturb the rapid accretion onto the protostar.
\par
The middle and right columns of Figure~\ref{fig:Snapshot} show the snapshots at the end of the simulation when $M_*=11.8~\Msun$.
While the gas structure around the protostar is complicated due to the turbulence, the protostar rotates and has a flattened shape. This is because the accreting gas brings into the protostar the angular momentum of the initial turbulence.
A circumstellar disk is about to emerge at this time.
In our current definition of the stellar surface, which compares the radial velocity to the sound velocity, the white ellipse approximates a protostar as shown in Figure~\ref{fig:Snapshot}.
Nevertheless, even inside this ellipse, it is the centrifugal force rather than the pressure gradient that is balanced by gravity in an outer part near the equatorial plane. From this point of view, such a part can be interpreted as a circumstellar disk.
The middle panels show the stellar interior is much hotter than the surrounding gas, which is nearly constant at several thousands of Kelvin with some fluctuations.
Moreover, as seen in the top and bottom panels, $L_\mathrm{iso}$ is anti-correlated with the density, indicating that the stellar radiation predominantly escapes through low-density regions.
\par
Figure \ref{fig:Turb_profile} shows the radial profiles of physical quantities at the end of the simulation.
As seen in panel (a), the equatorial density is basically larger than the polar one because of the rotation.
Panels (b) and (c) show that the radial infall velocity abruptly decreases and the temperature rises after the shock front.
\par
Panel (d) shows similar $L_\mathrm{iso}$ profiles, regardless of the equatorial and polar directions. The luminosity is very low owing to the high optical depth in the deep stellar interior.
Outside the protostar, the luminosity is sufficiently large that, unlike the case with $M_*=0.2~\Msun$ in Figure~\ref{fig:Sphe_profile}, the emissivity of the infalling gas hardly enhances the luminosity.
\par
Panel (e) shows the spherically averaged angular velocity distribution. For comparison, the dashed line also shows the Keplerian angular velocity $\Omega_\mathrm{Kep}=\sqrt{GM_r/r}$, where $M_r$ is the enclosed mass within the radius $r$.
The angular velocity is almost constant in the stellar interior, which means rigid rotation. The protostar rotates rapidly at more than 0.4 times the Keplerian velocity at any given radius. Especially near the surface at $r \simeq 5$~AU, the rotational velocity is very close to the Keplerian value.

%
%
\section{Discussions} \label{Sec:Discussions}
We have developed new RHD solver and follow the protostellar evolution in SMS formation with 3D RHD simulation until its mass reaches $\sim10\Msun$.
Our simulations show a swollen protostar steadily grows in mass by accretion, which differs from the previous result by \citet{Luo_et_al_2018}. This may be attributed to differences in the RHD solver or initial conditions, but the main reason is currently uncertain and open for further studies.
Regarding the protostellar rotation, our results show the rigid rotation inside the protostar and nearly Keplerian motion near the surface. This is the same trend as in \citet{Greif_et_al_2012} and \citet{Stacy_et_al_2013}, who study the ordinary Pop III star formation without solving the radiative transfer. They found that protostars forming in various mini-halos generally have high spins.
Their and our results suggest that, if there are no processes efficiently extracting the angular momentum, the primordial stars should be rapid rotators. Their structure and evolution should differ from the non-rotating case.
\par
Although we have terminated the simulations when $M_* \sim 10~\Msun$, we can follow a longer-term evolution 
with our code, if we implement nuclear fusion processes effective in the later stage ($M\gtrsim50~\Msun$).
This will reveal directly how the protostar evolves under the strong rotation and whether it grows into a SMS as predicted by 1D stellar evolution calculations. The stellar spin affects the evolution in various ways.
First, it potentially triggers some instabilities which generate turbulence. The turbulence induces chemical mixing, and it may change the nuclear fusion rate and the overall stellar structure. Second, the rotation stabilizes the SMSs and raises their maximum masses by an order of magnitude compared to the non-rotating cases \citep{Haemmerle_2021}.
The masses of the seed BHs can be consequently elevated, which assists the formation of the SMBHs observed at high redshifts.
\par
A circumstellar disk forms at the end of our simulation, and it should continue to grow in mass and size afterward. The disk accretion process onto a rapidly rotating SMS is still under debate.
For instance, \citet{Lee+Yoon2016} argue that the $\Omega\Gamma$ limit regulates the gas accretion. \citet{Takahashi_and_Omukai_2017} show that the steady disk accretion solution for an arbitrary angular momentum flux exists, concluding that protostars can grow even under the $\Omega\Gamma$ limit. Whereas these authors only rely on 1D models, our 3D RHD simulations can reveal the realistic disk structure and the accretion process.
The circumstellar disk is also expected to become gravitationally unstable and fragment \citep{Becerra_et_al_2015, Kimura_et_al_2021}. If some of the emerging fragments fall onto the central rotating protostar, it may alter the protostellar evolution with resulting violently variable accretion rates \citep{Sakurai16}.
Further 3D RHD simulations studying these effects are necessary for elucidating the evolution of the star-disk system.
\section{Summary} \label{Sec:Summary}
In this paper, we have studied the protostellar evolution under very rapid accretion supposed for the DC, performing 3D RHD simulations resolving the stellar interior structure.
We have used the AMR code SFUMATO-RT, in which we have implemented a newly-developed RHD solver employing an explicit M1 closure method. We follow the evolution for about 10~yr after the protostar formation, during which the stellar mass increases to $\sim 10~\Msun$.
\par
In one case starting from a spherical cloud, a swollen protostar forms as expected by previous 1D stellar evolution calculations.
At the end of the simulation when $M_* \sim 10~\Msun$, the stellar radius is $\simeq 1000~\Rsun$, the luminosity is $\sim10^5~\Lsun$, and the effective temperature is $\simeq 4000$~K.
Such protostars with the cool atmosphere emit a negligible amount of ionizing photons, and the resulting radiative feedback is too weak to disturb the accretion.
\par
In the other case starting from a turbulent cloud,
a rapidly-rotating swollen protostar appears with some different properties from the counterpart in the spherical case.
The protostar rotates at more than about $0.4$ times the Keplerian velocity everywhere in the interior due to the angular momentum brought by the accreting gas.
As a result, the equatorial radius is more than twice as large as the polar one. Also in this case, the effective temperature remains several thousand Kelvin and the radiative feedback is ineffective.
\par
We have followed the protostellar birth and subsequent evolution for $\sim 10$~yr expected in the DC. This is only the beginning of the evolution until the seed BH formation after $\sim$~Myr. Our 3D RHD simulations have a potential to reveal the realistic protostellar evolution in the later stage, for which only 1D modeling has been applied so far.
%
%
\acknowledgments
We are grateful to Tomoaki Matsumoto, Kazuyuki Omukai, Kunihito Ioka, Ryoki Matsukoba, Shinsuke Takasao, Kengo Tomida and Takahiro Tanaka for fruitful discussions and comments.
The numerical simulations were carried out on XC50  {\tt Aterui II} at the Center for Computational Astrophysics (CfCA) of the National Astronomical Observatory of Japan, and Yukawa-21 at Yukawa Institute for Theoretical Physics of Kyoto University.
This research could never be accomplished without the support by Grants-in-Aid for Scientific Research (TH:19H01934, 21H00041; KS: 21K20373) from the Japan Society for the Promotion of Science. This work is also supported by JST SPRING, Grant Number JPMJSP2110 (KK), the ANRI Fellowship (KK), and the Hakubi Project Funding of Kyoto University (KS).

\end{document}